\documentclass{article}%
\usepackage{amsmath}
\usepackage{hyperref}
\usepackage{amsfonts}
\usepackage{amssymb}
\usepackage{graphicx}%
\begin{document}

\title{Analytic Scalar Field Theory}
\author{Thomas Curtright$^{\S }$ and Galib Hoq$^{\flat}$\medskip\\Department of Physics, University of Miami, Coral Gables, FL 33124\medskip
\and $^{\S }$ curtright@miami.edu\ \ \ \ \ $^{\flat}$ gxh368@miami.edu}
\date{}
\maketitle

\begin{abstract}
We discuss scalar field theories with potentials $V\left(  \phi\right)
=\kappa\left(  \phi^{2}\right)  ^{\nu}$ for generic $\nu$. \ We conjecture
that these models evade various no-go theorems for scalar field theories in
four spacetime dimensions.

\end{abstract}

In four-dimensional spacetime (4D) the two simplest renormalizable scalar
field theories are well-known to be flawed. \ Consider the standard Lagrangian
density
\begin{equation}
L=\frac{1}{2}\left(  \partial\phi\right)  ^{2}-\frac{1}{2}m^{2}\phi
^{2}-V\left(  \phi\right)
\end{equation}
If $V\left(  \phi\right)  =\kappa\phi^{3}$, the theory has no ground
state,\ i.e. the energy density is unbounded below, as is evident from
consideration of semi-classical coherent states \cite{Baym}. \ If $V\left(
\phi\right)  =\kappa\phi^{4}$ for real coupling constant $\kappa>0$, the
theory has a ground state, by the same considerations, but it is
\textquotedblleft trivial\textquotedblright\ due to renormalization effects in
4D, i.e. only $\kappa=0$ is sensible under renormalization
\cite{Frolich,AD,Kopper}.

Moreover, if $V$ involves higher integer powers of $\phi$ in 4D the theory is
non-renormalizable by well-known power counting arguments. \ As a consequence
of all this the general folklore is: \ \emph{In 4D there are no consistent and
tractable interacting quantum field theories involving only a scalar field.}

Nevertheless, around 60 years ago there were intriguing proposals, by Efimov
\cite{Efimov}, by Fradkin \cite{Fradkin}, and by Delbourgo, Salam, and
Strathdee \cite{Salam}, that non-polynomial interactions could perhaps lead to
interesting and non-trivial renormalizable quantum field theories involving
only scalar fields, even in 4D, albeit at considerable computational expense.
\ More recently the aforementioned general folklore has also been called into
question by proponents of PT-symmetric theories \cite{PT}, but again at
considerable computational expense.

As argued here, the underlying issues and their possible work-arounds can be
considered in a tractable framework through the use of specific analytic
functions of $\phi$, with computations more manageable than those encountered
in previous work.\ \ The models proposed here are less challenging than the
generic situations studied in \cite{Salam}, while PT-symmetric systems are
essentially a special case after only minor modifications.

Here it is proposed that there exists a well-defined theory with $V\left(
\phi\right)  =\kappa\left(  \phi^{2}\right)  ^{\nu}$ for generic real
$\nu>-1/2$, where the positive principal $\nu$th root is understood. \ Upon
considering real coherent state field configurations for which $\phi^{2}>0$,
this model has a positive energy density, at least for positive real $\kappa$.
\ In addition, this model admits a non-trivial but manageable perturbative
analysis in powers of $\kappa$, even in 4D for all real $\nu<2$. \ There is a
previous study of this model that relies on perturbative expansions in both
$\kappa$ and the exponent $\nu$ \cite{prePT}. \ In contrast to that earlier
study, the new method presented here routinely gives exact results analytic in
$\nu$ for each fixed order in $\kappa$. \ 

To deal with an arbitrary power of $\phi^{2}$ our novel approach is to define
$V\left(  \phi\right)  =\kappa\left(  \phi^{2}\right)  ^{\nu}$ as \emph{a
linear superposition of Gaussians}. \ That is to say, we use the exact
mathematical result%
\begin{equation}
\left(  \phi^{2}\right)  ^{\nu}=\Gamma\left(  1+\nu\right)  \frac{i}{2\pi}%
{\displaystyle\oint}
\left(  -z\right)  ^{-1-\nu}\exp\left(  -z\phi^{2}\right)  dz\label{BasicID}%
\end{equation}
where $\phi^{2}$\ is assumed to be real and positive, and where the
integration $%
{\displaystyle\oint}
$\ is along \href{https://en.wikipedia.org/wiki/Hankel_contour}{a Hankel
contour} around the positive $z$ half-line \cite{Hankel,f(R)}. \ The
right-hand side of (\ref{BasicID}) is well-defined and completely unambiguous
so long as $\nu$ is not a negative integer. \ In particular, a
\textquotedblleft marginal\textquotedblright\ interaction in $N$ spacetime
dimensions (ND) is given by $\nu=N/\left(  N-2\right)  $. \ In ND the coupling
constant $\kappa$ is always dimensionless for potential energy density
$V\left(  \phi\right)  =\kappa\left(  \phi^{2}\right)  ^{\frac{N}{N-2}}$.

As an illustration of this new method, consider the $O\left(  \kappa\right)  $
contribution to the vacuum expectation value (VEV) of the potential energy
density. \ To this order in $\kappa$, only the free-field VEV of a Gaussian is
required to complete the calculation. \ As shown in the Appendix this is%
\begin{equation}
\left\langle \exp\left(  -z\phi^{2}\right)  \right\rangle =\frac{1}%
{\sqrt{1+2z\left\langle \phi^{2}\right\rangle }} \label{OneGaussianVEV}%
\end{equation}
The calculation therefore reduces to the free-field evaluation of
$\left\langle \phi^{2}\right\rangle $, which can be done in ND using a
momentum cut-off, or $\zeta$ function regularization, or dimensional
regularization. \ All that remains is evaluation of the manifestly convergent
contour integral in (\ref{BasicID}). \ Thus, by rescaling the variable of
integration,%
\begin{equation}
\left\langle \left(  \phi^{2}\right)  ^{\nu}\right\rangle =\Gamma\left(
1+\nu\right)  \frac{i}{2\pi}%
{\displaystyle\oint}
\left(  -z\right)  ^{-1-\nu}\frac{1}{\sqrt{1+2z\left\langle \phi
^{2}\right\rangle }}dz=\left\langle \phi^{2}\right\rangle ^{\nu}\Gamma\left(
1+\nu\right)  \frac{i}{2\pi}%
{\displaystyle\oint}
\left(  -z\right)  ^{-1-\nu}\frac{1}{\sqrt{1+2z}}dz
\end{equation}
where $\left\langle \phi^{2}\right\rangle $\ is assumed to be positive. \ The
integral yields to routine contour evaluation methods that take into account
the analytic structure of the integrand. \ The result is%
\begin{equation}
\frac{i}{2\pi}%
{\displaystyle\oint}
\left(  -z\right)  ^{-1-\nu}\frac{1}{\sqrt{1+2z}}dz=\frac{2^{\nu}}{\sqrt{\pi}%
}\frac{\Gamma\left(  \nu+1/2\right)  }{\Gamma\left(  \nu+1\right)  }%
\end{equation}
The final result for the VEV in question is then%
\begin{equation}
\left\langle V\left(  \phi\right)  \right\rangle =\kappa\left\langle \left(
\phi^{2}\right)  ^{\nu}\right\rangle =\frac{2^{\nu}}{\sqrt{\pi}}~\Gamma\left(
\nu+1/2\right)  ~\kappa\left\langle \phi^{2}\right\rangle ^{\nu}
\label{PotentialVEV}%
\end{equation}
The coefficient in this result is just an analytic continuation of the double
factorial for odd integers, $\left(  2\nu-1\right)  !!$.

So, at least to this order in $\kappa$, quantum effects have only modified the
form expected from classical field theory by a simple replacement of $\kappa$
with
\begin{equation}
\kappa_{\operatorname*{effective}}=2^{\nu}\Gamma\left(  \nu+1/2\right)
~\kappa/\sqrt{\pi} \label{KappaEffective}%
\end{equation}
where all the $\nu$ dependence is \emph{exact} and explicit. \ Note that
(\ref{PotentialVEV}) is valid even for some negative powers of $\phi^{2}$, at
least so long as $\nu>-1/2$ if the coefficient of the VEV is to be finite and positive.

Similar results follow for the $O\left(  \kappa\right)  $ corrections to the
scalar propagator, which are again \emph{exact} as functions of $\nu$, as can
be obtained just by differentiation of $\left\langle V\left(  \phi\right)
\right\rangle $. \ To this order the effect is only a shift in the mass of the
field as is evident upon summing the usual geometric series involving the
$O\left(  \kappa\right)  $ correction.%
\begin{equation}
\Delta\left(  k\right)  =\frac{i}{k^{2}-m^{2}-2\nu\kappa
_{\operatorname*{effective}}\left\langle \phi^{2}\right\rangle ^{\nu-1}}%
\end{equation}
Momentum dependent corrections to $\Delta\left(  k\right)  $\ begin at order
$\kappa^{2}$. \ These will be discussed elsewhere \cite{Deep}. \ However, the
method to compute higher order corrections is a straightforward procedure
based on (\ref{BasicID}).

For example, the order $\kappa^{2}$ computation of $\left\langle V\left(
\phi\right)  \right\rangle $ follows from $\left\langle \left(  \phi
^{2}\left(  0\right)  \right)  ^{\nu}\left(  \phi^{2}\left(  x\right)
\right)  ^{\nu}\right\rangle $ upon replacing the powers of $\phi^{2}$ with
superpositions of Gaussians as defined in (\ref{BasicID}). \ Evaluation of the
VEV is then facilitated by the identity%
\begin{equation}
\left\langle e^{-w\phi^{2}\left(  0\right)  }e^{-z\phi^{2}\left(  x\right)
}\right\rangle =\frac{1}{\sqrt{1+\left(  2w+2z\right)  \left\langle \phi
^{2}\left(  0\right)  \right\rangle +4wz\left(  \left\langle \phi^{2}\left(
0\right)  \right\rangle ^{2}-\left\langle \phi\left(  x\right)  \phi\left(
0\right)  \right\rangle ^{2}\right)  }} \label{TwoGaussianVEV}%
\end{equation}
This identity, as well as that for higher multinomials involving Gaussians,
follows from a
\href{https://en.wikipedia.org/wiki/Hubbard-Stratonovich_transformation}{Stratonovich-Hubbard}
transformation \cite{Strat,Hub}, or else from standard
\href{https://en.wikipedia.org/wiki/Wick's_theorem}{Wick contraction
combinatorics} \cite{Wick}, as given by%
\begin{equation}
\left\langle e^{-w\phi^{2}\left(  0\right)  }\phi^{2n}\left(  x\right)
\right\rangle =\frac{\left\langle \phi^{2}\left(  0\right)  \right\rangle
^{n}}{\left(  1+2w\left\langle \phi^{2}\left(  0\right)  \right\rangle
\right)  ^{1/2}}\sum_{k=0}^{n}\dbinom{2n}{2k}\frac{\left(  2n-2k-1\right)
!!\left(  2k-1\right)  !!}{\left(  1+2w\left\langle \phi^{2}\left(  0\right)
\right\rangle \right)  ^{k}}\left(  -2w\frac{\left\langle \phi\left(
x\right)  \phi\left(  0\right)  \right\rangle ^{2}}{\left\langle \phi
^{2}\left(  0\right)  \right\rangle }\right)  ^{k}%
\end{equation}
The generating functional for the latter VEVs is then obtained by taking
$\sum_{n=0}^{\infty}\frac{\left(  -z\right)  ^{n}}{n!}\phi^{2n}\left(
x\right)  $ to obtain (\ref{TwoGaussianVEV}). \ 

For this example the integration over Hankel contours once again yields to
routine evaluation methods. The final result is a
\href{https://en.wikipedia.org/wiki/Hypergeometric_function}{Gauss
hypergeometric function}.
\begin{equation}
\kappa^{2}\left\langle \left(  \phi^{2}\left(  0\right)  \right)  ^{\nu
}\left(  \phi^{2}\left(  x\right)  \right)  ^{\nu}\right\rangle =\kappa
_{\text{effective}}^{2}\left\langle \phi^{2}\left(  0\right)  \right\rangle
^{2\nu}\left.  _{2}F_{1}\left(  -\nu,-\nu;~\frac{1}{2};~\frac{\left\langle
\phi\left(  x\right)  \phi\left(  0\right)  \right\rangle ^{2}}{\left\langle
\phi^{2}\left(  0\right)  \right\rangle ^{2}}\right)  \right.
\end{equation}
where again $\kappa_{\text{effective}}$ is given by (\ref{KappaEffective}).
\ It is satisfying that this result behaves in accord with the
\href{https://en.wikipedia.org/wiki/Cluster_decomposition}{cluster
decomposition} property \cite{Cluster}\ in the sense that $\kappa
^{2}\left\langle \left(  \phi^{2}\left(  0\right)  \right)  ^{\nu}\left(
\phi^{2}\left(  x\right)  \right)  ^{\nu}\right\rangle \underset{\left\langle
\phi\left(  x\right)  \phi\left(  0\right)  \right\rangle \rightarrow
0}{\longrightarrow}\left(  \kappa\left\langle \left(  \phi^{2}\left(
0\right)  \right)  ^{\nu}\right\rangle \right)  ^{2}$.

A special case of interest is given by%
\begin{equation}
\kappa^{2}\left\langle \left(  \phi^{2}\left(  0\right)  \right)
^{3/2}\left(  \phi^{2}\left(  x\right)  \right)  ^{3/2}\right\rangle =\frac
{8}{\pi}\kappa^{2}\left\langle \phi^{2}\left(  0\right)  \right\rangle
^{3}\left.  _{2}F_{1}\left(  -\frac{3}{2},-\frac{3}{2};~\frac{1}{2}%
;~\frac{\left\langle \phi\left(  x\right)  \phi\left(  0\right)  \right\rangle
^{2}}{\left\langle \phi^{2}\left(  0\right)  \right\rangle ^{2}}\right)
\right.  \label{PhiSquaredThreeHalves}%
\end{equation}
where this particular $\left.  _{2}F_{1}\right.  $ reduces to elementary
functions.%
\begin{align}
\left.  _{2}F_{1}\left(  -\frac{3}{2},-\frac{3}{2};~\frac{1}{2};~z^{2}\right)
\right.   &  =\frac{1}{4}\left(  \left(  4+11z^{2}\right)  \sqrt{1-z^{2}%
}+3\left(  3+2z^{2}\right)  z\arcsin z\right) \nonumber\\
&  =1+\frac{9}{2}z^{2}+\frac{3}{8}z^{4}+\frac{1}{80}z^{6}+\frac{9}{4480}%
z^{8}+O\left(  z^{10}\right)
\end{align}
Compare (\ref{PhiSquaredThreeHalves}) to that of conventional $\kappa\phi^{3}$
theory \cite{Baym} for which there are only two terms in the corresponding
VEV, rather than an infinite series, namely, \
\begin{equation}
\kappa^{2}\left\langle \phi\left(  0\right)  ^{3}\phi\left(  x\right)
^{3}\right\rangle =\kappa^{2}\left(  9\left\langle \phi^{2}\left(  0\right)
\right\rangle ^{2}\left\langle \phi\left(  x\right)  \phi\left(  0\right)
\right\rangle +6\left\langle \phi\left(  x\right)  \phi\left(  0\right)
\right\rangle ^{3}\right)
\end{equation}
The comparison highlights the difference between $V=\kappa\phi^{3}$ and
$V=\kappa\left(  \phi^{2}\right)  ^{3/2}$ as defined by (\ref{BasicID}).

The above results raise the possibility of scalar field theories with non-zero
$\left\langle \phi^{2}\right\rangle $ through more exotic mechanisms than
usually encountered in the literature. \ Consider a\ potential defined as%
\begin{equation}
V\left(  \phi\right)  =\frac{\kappa_{1}}{\left(  \phi^{2}\right)  ^{\nu_{1}}%
}+\kappa_{2}\left(  \phi^{2}\right)  ^{\nu_{2}} \label{VComposite}%
\end{equation}
with real and positive $\kappa_{1}$, $\kappa_{2}$, $\nu_{1}$, \& $\nu_{2}$.
\ With the additional assumption $0<\nu_{1}<1/2$, free-field VEVs of both
terms in $V$ give an expectation value to first order in either $\kappa$ as%
\begin{equation}
\left\langle V\left(  \phi\right)  \right\rangle =\frac{\kappa_{1\text{
}\operatorname*{effective}}}{\left\langle \phi^{2}\right\rangle ^{\nu_{1}}%
}+\kappa_{2\text{ }\operatorname*{effective}}\left\langle \phi^{2}%
\right\rangle ^{\nu_{2}}%
\end{equation}
where the effective\ $\kappa_{1}$ and $\kappa_{2}$ are given by%
\begin{equation}
\kappa_{1\text{ }\operatorname*{effective}}=\frac{2^{\nu_{1}}\Gamma\left(
1-2\nu_{1}\right)  }{\Gamma\left(  1-\nu_{1}\right)  }~\kappa_{1}%
\ ,\ \ \ \kappa_{2\text{ }\operatorname*{effective}}=\frac{\Gamma\left(
1+2\nu_{2}\right)  }{2^{\nu_{2}}\Gamma\left(  1+\nu_{2}\right)  }~\kappa_{2}%
\end{equation}
The following are plots of the effective $\kappa$s as functions of their
respective $\nu$s, along with an example of a classical $V$ versus $\phi$ as
well as lowest-order $\left\langle V\right\rangle $ versus $\sqrt{\left\langle
\phi^{2}\right\rangle }$, for $\kappa_{1}=\kappa_{2}=1$, $\nu_{1}=1/3$, \&
$\nu_{2}=3/2$.

\noindent\ \ \ \ \ \ \ \ \ \
{\parbox[b]{2.6152in}{\begin{center}
\includegraphics[
height=1.6336in,
width=2.6152in
]%
{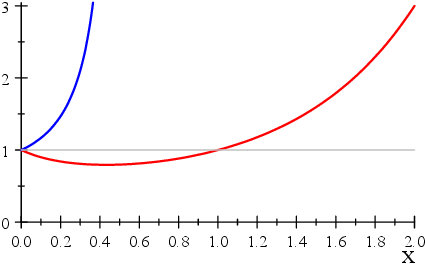}%
\\
$\kappa_{1\text{ }\operatorname*{effective}}/\kappa_{1}$ in blue and
$\kappa_{2\text{ }\operatorname*{effective}}/\kappa_{2}$ in red, as functions
of $x=\nu_{1}$ and $\nu_{2}$.
\end{center}}}
\ \ \ \ \ \ \ \ \
{\parbox[b]{2.6195in}{\begin{center}
\includegraphics[
trim=0.000000in 0.000000in -0.118178in 0.000000in,
height=1.6336in,
width=2.6195in
]%
{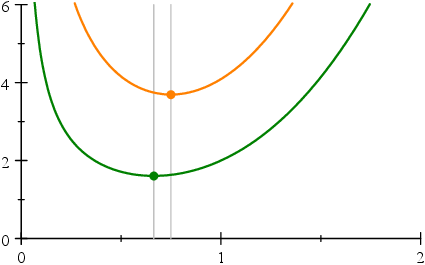}%
\\
$V\left(  \phi\right)  $, in green, \& $\left\langle V\right\rangle $ versus
$\sqrt{\left\langle \phi^{2}\right\rangle }$, in orange, with minima as
indicated.
\end{center}}}

\noindent Several parallels to the earlier work of Salam et al. \cite{Salam}
would result from power series expansions in $\phi$ about either of the minima shown.

The interacting models, as defined here, are in one respect similar to that
for a 2D sine-Gordon model \cite{Coleman} with $V\left(  \phi\right)
=\kappa\cos\left(  \phi\right)  $, where $\kappa$ has mass dimension two.
\ That is to say, perturbative results to any finite order in $\kappa$
\cite{Caveat} involve closed $n$-loop Feynman diagrams for \emph{all} integer
$n\geq1$. \ However, in contrast to the 2D sine-Gordon model for which $\phi$
has mass dimension $\left[  \phi\right]  =0$, if the models considered here
are to be in $N$ spacetime dimensions, then $\left[  \phi\right]  =\left(
N-2\right)  /2$. \ Hence the coupling constant for $V\left(  \phi\right)
=\kappa\left(  \phi^{2}\right)  ^{\nu}$ has mass dimension $\left[
\kappa\right]  =2\nu+\left(  1-\nu\right)  N$. \ Conventional power-counting
arguments, taken naively at face value, would suggest the model is
super-renormalizable for $\left[  \kappa\right]  >0$, renormalizable for
$\left[  \kappa\right]  =0$, and non-renormalizable for $\left[
\kappa\right]  <0$. \ To see whether these naive suggestions are correct is
one objective of future studies \cite{Deep}.

\subsection*{Appendix: \ Some combinatorics}

The free field VEV of a Gaussian follows from
\begin{equation}
\left\langle \exp\left(  -z\phi^{2}\right)  \right\rangle =\sum_{n=0}^{\infty
}\frac{\left(  -z\right)  ^{n}}{n!}\left\langle \phi^{2n}\right\rangle
\tag{A1}%
\end{equation}
where $\left\langle \phi^{2n}\right\rangle $ is a \textquotedblleft
flower\textquotedblright\ graph with $n$ \textquotedblleft
petals\textquotedblright\ as sketched below, each petal representing the VEV
$\left\langle \phi^{2}\right\rangle $.%
\begin{center}
\includegraphics[
height=2.7466in,
width=2.8746in
]%
{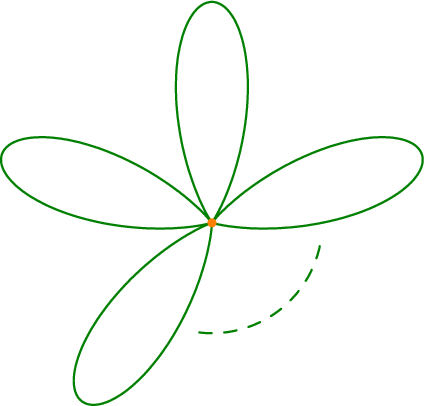}%
\\
A Feynman flower diagram.
\end{center}
The combinatoric factor for this graph is $\left(  2n-1\right)  !!=\left(
2n\right)  !/\left(  2^{n}n!\right)  $ where $\left(  2n\right)  !$
corresponds to the number of ways to select each individual $\phi$ from
$\phi^{2n}$, each petal has a symmetry factor of $1/2$, and the additional
$1/n!$ is a symmetry factor that corresponds to the number of permutations of
the petals. \ That is to say,%
\begin{equation}
\left\langle \phi^{2n}\right\rangle =\left(  2n-1\right)  !!\left\langle
\phi^{2}\right\rangle ^{n} \tag{A2}%
\end{equation}
where, e.g., after Wick rotation using dimensional regularization in $N$
Euclidean dimensions,
\begin{equation}
\left\langle \phi^{2}\right\rangle =\int\frac{1}{k^{2}+m^{2}}~d^{N}k\equiv
m^{N-2}\pi^{N/2}\Gamma\left(  1-N/2\right)  \tag{A3}%
\end{equation}
This immediately gives the result (\ref{OneGaussianVEV}) in the text, namely,
\begin{equation}
\left\langle \exp\left(  -z\phi^{2}\right)  \right\rangle =\sum_{n=0}^{\infty
}\frac{\left(  2n-1\right)  !!}{n!}\left(  -z\left\langle \phi^{2}%
\right\rangle \right)  ^{n}=\frac{1}{\sqrt{1+2z\left\langle \phi
^{2}\right\rangle }} \tag{A4}%
\end{equation}
This result also follows from standard operator methods, e.g., as often used
to prove
\href{https://en.wikipedia.org/wiki/Baker-Campbell-Hausdorff_formula}{the BCH
formula}.\bigskip

As explained in the text following (\ref{TwoGaussianVEV}), similar
combinatorics, as well as other methods, give the result for the VEV of two or
more Gaussians.

\end{document}